\documentclass{sig-alternate-hotpets}
\usepackage{graphicx}
\usepackage{subfig}
\usepackage{hyperref}       
\usepackage{url} 
\begin{document}
\title{ML Privacy Meter: Aiding Regulatory Compliance by Quantifying the Privacy Risks of Machine Learning\thanks{Repository for the code and tutorials is available at \url{https://github.com/privacytrustlab/ml_privacy_meter}}}

\author{Sasi Kumar Murakonda, Reza Shokri \\ \\ \affaddr{Data Privacy and Trustworthy ML Research Lab} \\ \affaddr{National University of Singapore} \\ \email{\{murakond,reza\}@comp.nus.edu.sg}}

\maketitle

\begin{abstract}

When building machine learning models using sensitive data, organizations should ensure that the data processed in such systems is adequately protected. For projects involving machine learning on personal data, Article 35 of the GDPR mandates it to perform a Data Protection Impact Assessment (DPIA). In addition to the threats of illegitimate access to data through security breaches, machine learning models pose an additional privacy risk to the data by indirectly revealing about it through the model predictions and parameters. Guidances released by the Information Commissioner’s Office (UK) and the National Institute of Standards and Technology (US) emphasize on the threats to data from models and recommend organizations to account for and estimate these risks to comply with data protection regulations. Hence, there is an immediate need for a tool that can quantify the privacy risks to data from models.

In this paper, we focus on this indirect leakage about training data from machine learning models. We present ML Privacy Meter, a tool that can quantify the privacy risk to data from models through state of the art membership inference attack techniques. We discuss how this tool can help practitioners in compliance with data protection regulations, when deploying machine learning models.

\end{abstract}

% !TEX root = ../main.tex
% \section{Privacy Risks of Machine Learning}
\section{Data Privacy Risks of Machine \\ Learning Models}
\label{sec:intro}

% \textbf{Point 1:} Data collection

Organizations are collecting massive amounts of personal information for building applications that are powered by machine learning. This data, which is used to train the models, typically contain sensitive information about individuals. Machine learning models encode information about the datasets on which they are trained. The encoded information is supposed to reflect the general patterns underlying the population data. However, it is commonly observed that these models memorize specific information about some members of their training data~\cite{carlini2019secret} or be tricked to do so~\cite{song2017machine}.

Models with high generalization gap as well as the models with high capacity (such as deep neural networks) are more susceptible to memorizing data points from their training set. This is reflected in the predictions of the model, which exhibits a different behavior on training data versus test data, and in the model's parameters which store statistically correlated information about specific data points in their training set~\cite{shokri2017membership,nasr2019comprehensive}. This vulnerability of machine learning models was shown using membership inference attacks, where an attacker detects the presence of a particular record in the training dataset of a model, just by observing the model. Machine learning models were shown to be susceptible to these attacks in both the black-box~\cite{shokri2017membership} and white-box settings~\cite{nasr2019comprehensive}.

In the black-box setting, we can only observe predictions of the model. This setting models the scenario of machine learning as a service offered on cloud platforms by companies such as Amazon,\footnote{\url{https://aws.amazon.com/machine-learning}} Microsoft,\footnote{\url{https://studio.azureml.net}} and Google.\footnote{\url{https://cloud.google.com/prediction}}. It can be used to measure the privacy risks against legitimate users of a model who seek predictions on their queries. In the white-box setting, we can also observe the parameters of the model. This reflects the scenario where a model is outsourced to a potentially untrusted server or to the cloud, or is shared with an aggregator in the federated learning setting~\cite{mcmahan2017communication,shokri2015privacy}. The privacy risks of machine learning models can be evaluated as the accuracy of such inference attacks against their training data.  

\section{Data Protection Regulations}

For a safe and secure use of machine learning models, it is important to have a quantitative assessment of the privacy risks of these models, and to make sure that they do not reveal sensitive information about their training data. Data protection regulations, such as GDPR, and AI governance frameworks require personal data to be protected when used in AI systems, and that the users have control over their data and awareness about how it is being used. 

For projects involving innovative technologies such as machine learning, it is mandatory from Article 35 of the GDPR to perform a Data Protection Impact Assessment (DPIA).\footnote{\url{https://gdpr-info.eu/art-35-gdpr/}} The key steps in DPIA are to identify the potential threats to data and assess how they might affect individuals. In general, risk assessment in DPIA statements focuses on the risk of security breaches and illegitimate access to the data. Machine learning models pose additional privacy risk to the training data by indirectly revealing about it through the model's predictions and parameters. Hence, special attention needs to be paid for data protection rules in AI regulation frameworks. Guidances released by both the European Commission and the White House call for protection of personal data during all the phases of deploying AI systems and build systems that are resistant to attacks~\cite{EU-AI,WH-AI}. Recent reports published by the Information Commissioner’s Office (ICO) for auditing AI~\cite{ICO} and the National Institute of Standards and Technology (NIST) for securing applications of Artificial Intelligence~\cite{NIST} highlight the privacy risk to data from machine learning models. And they specifically mention membership inference as a confidentiality violation and potential threat to the training data from models. It is recommended in the auditing framework by ICO for organizations to identify these threats and take measures to minimize the risk~\cite{ICO}. As the ICO’s investigation teams will be using this framework to assess the compliance with data protection laws, organizations must account for and estimate the privacy risks to data through models. 

% 1. Both ICO and NIST say there is privacy risk to data from models and explicitly mention membership inference attacks as potential threat.

% 2. ICO even went to the point of saying such risks should be included in DPIA for compliance. 

\section{ML Privacy Meter}
% \textbf{Point 6:} Practical scenario

A tool that can automatically assess the privacy risks of machine learning models to their training data can aid practitioners in compliance with data protection regulations. But how do we measure the risk of indirect information leakage about training data from complex ML models? We present ML Privacy Meter that can quantify the privacy risks to training data and is based on well-established algorithms to measure privacy risks of machine learning models through membership inference attacks \cite{nasr2019comprehensive,shokri2017membership}. The tool provides privacy risk scores that help in identifying the data records that are under high risk of being revealed through the model parameters or predictions. The tool can generate extensive privacy reports about the aggregate and individual risk for data records in the training set at multiple levels of access to the model. It can estimate the amount of information that can be revealed through the predictions of a model (referred to as Black-box access) and through both the predictions and parameters of a model (referred to as White-box access). Hence, when providing query access to the model or revealing the entire model, the tool can be used to assess the potential threats to training data.

\begin{figure}[t]
\centering
\caption{\small{ML Privacy Meter is a python library that enables quantifying the privacy risks of machine learning models to members in the training dataset. The tool provides privacy risk scores which help in identifying the data records that are under high risk of being revealed through the model parameters or predictions.}}
\label{fig:ML Privacy Meter}
\includegraphics[width=0.5\textwidth]{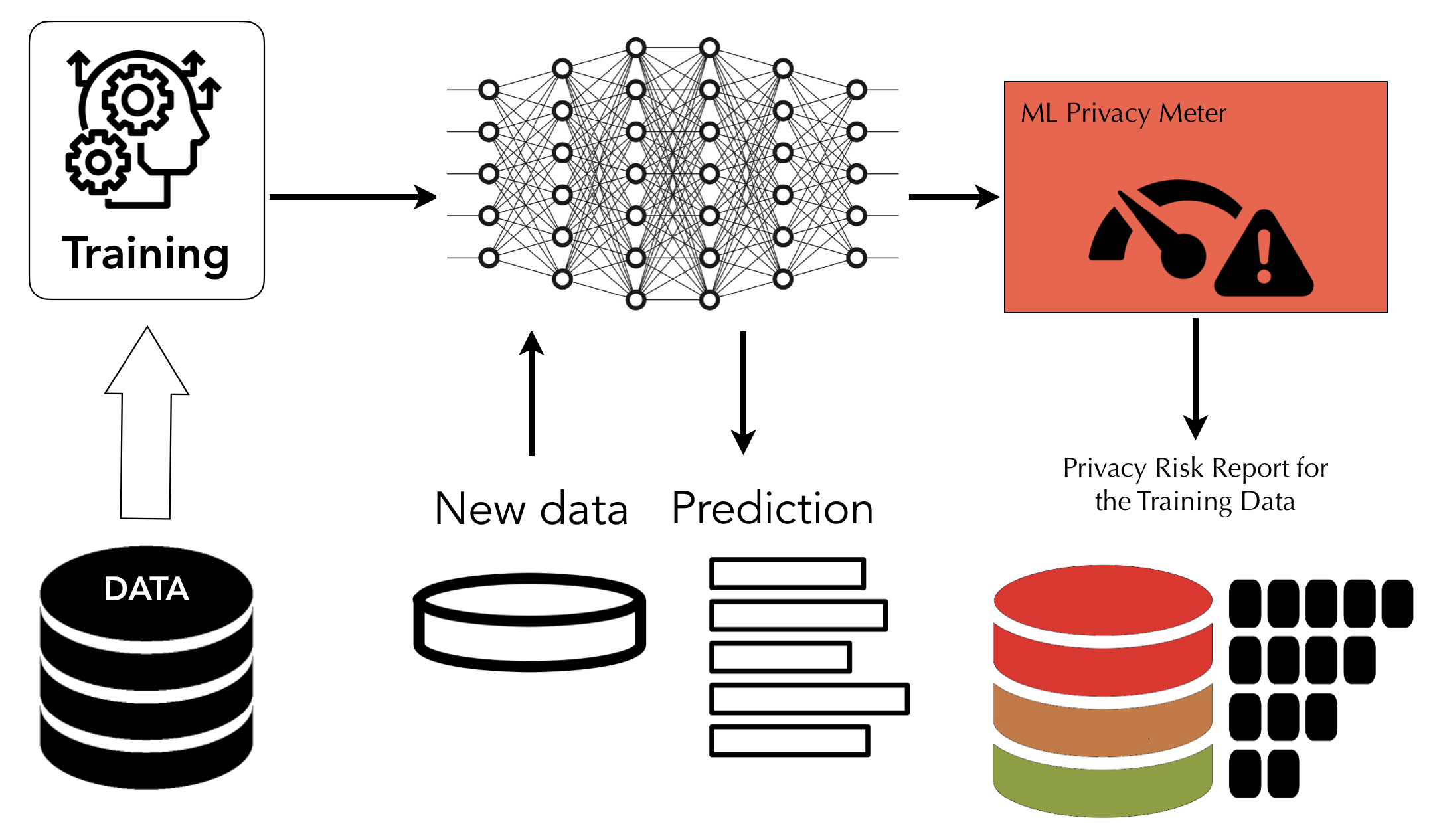}
\end{figure}

\begin{figure}[t]
\centering
\caption{\small{ML Privacy Meter quantifies the privacy risk to training data from machine learning models. The risk is measured through success of membership inference attacks quantified by an ROC curve representing the trade-off between true positive and false positive rates. It also allows for comparison of privacy risk across records from different classes.}}
\label{fig:plots}
\subfloat{\includegraphics[width=0.25\textwidth]{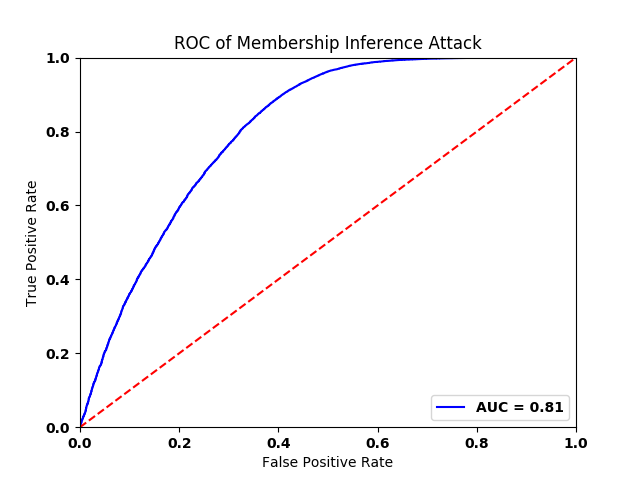}}
\subfloat{\includegraphics[width=0.25\textwidth]{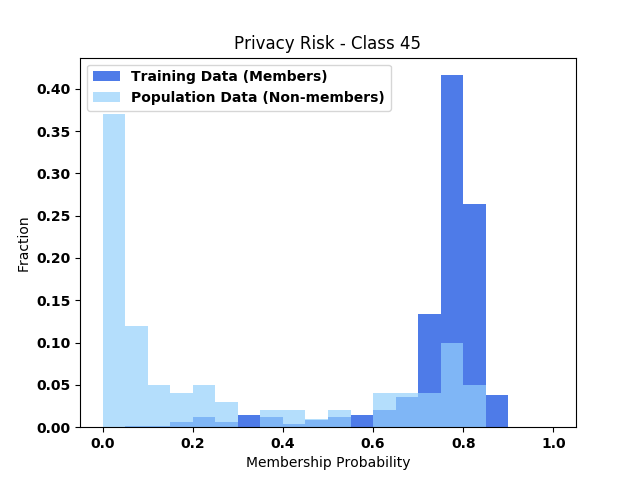}}
\end{figure}

ML Privacy Meter works by implementing membership inference attacks against machine learning models. It simulates attackers with different levels of access and knowledge about the model. It considers attackers who can exploit only the predictions of the model, the loss values, and the parameters of the model. For each of the simulated attacks, the tool reports risk scores for all the data records. These scores represent the attacker’s belief that the record was part of the training dataset. The larger the gap between the distribution of these scores for records that are in the training set versus records that are not in the training set, the larger is the leakage from the model would be. 

Success of the attacker can be quantified by an ROC curve representing the trade-off between False Positive Rate and True Positive Rate of the attacker. True positive represents correctly identifying a member as present in the data and False positive refers to identifying a non-member as member. An attack is successful if it can achieve larger values of True Positive rate at small values of False Positive rate. A trivial attack such as random guess can achieve equal True Positive and False Positive Rates. ML Privacy Meter automatically plots the trade-offs that are achieved by our simulated attackers. The area under those curves quantifies the aggregate privacy risk to the data posed by the model. The higher the area under curve, larger the risk. These numbers not only quantify the success of membership inference attacks, but they can also be seen as a measure of information leakage from the model.

When deploying machine learning models, this quantification of risk can be useful while performing a Data Protection Impact Assessment. The aim of doing a DPIA is to analyze, identify and minimize the potential threats to data. ML privacy meter can guide practitioners in all the three steps. It can help in estimating the privacy risk to data and to identify the potential causes of this risk. It can also be useful in selecting and deploying appropriate risk mitigation measures. 

The tool produces detailed privacy reports for the training data. It allows comparing the risk across records from different classes in the data. We can also compare the risk posed by providing black box access to the model with the risk due to white box access. As the tool can immediately measure the privacy risks for training data, practitioners can take simple actions such as finetuning their regularization techniques, sub-sampling, re-sampling their data, etc., to reduce the privacy risk. Or they can even choose to learn with a privacy protection, such as differential privacy, in place.  

Differential Privacy is a cryptographic notion of privacy, wherein the outputs of a computation should be indistinguishable when any single record in the data is modified. The level of indistinguishability is controlled by a privacy parameter $\epsilon$. Open source tools such as OpenDP \footnote{\url{https://github.com/opendifferentialprivacy/}} and TensorFlow Privacy \footnote{\url{https://github.com/tensorflow/privacy}} are available for training models with differential privacy guarantees. Selecting an appropriate value for $\epsilon$ is highly non-trivial when using these tools. Models learned with smaller value of $\epsilon$ provide better privacy guarantees but are also less accurate. $\epsilon$ represents a worst case upper bound on the privacy risk and the practical risk might be much lower. ML Privacy Meter can help in the selection of privacy parameters ($\epsilon$) for differential privacy by quantifying the risk posed at each value of epsilon. Compared to just relying on the guarantees provided by epsilon, using this method helps in deploying models with higher accuracy. By letting practitioners choose models with better utility, ML Privacy Meter can enable the use of privacy risk minimization techniques.
 
% In general, the tool can be used for estimating the trade-off between privacy risk and utility of any privacy protection mechanism.  In general, the tool can be used for estimating the trade-off between privacy risk and utility of any privacy protection mechanism.

% \textbf{Point 10:} Final remarks
\section{Summary}

By leaking information through predictions and parameters, machine learning models pose an additional privacy risk to data in AI systems. To comply with data protection regulations, we need to assess these risks and take possible mitigation measures. ML Privacy Meter quantifies the privacy risk of machine learning models to their training data. It can guide practitioners in regulatory compliance by helping them analyze, identify, and minimize the threats to data. By permitting for deploying models with better accuracy, through practical estimates of utility-privacy trade-offs, we expect the tool to boost adaptation of privacy enhancing techniques in machine learning.

\bibliographystyle{abbrv}
% \bibliography{references}  

\end{document}